%% file: main.tex
\documentclass[journal,12pt,draftclsnofoot,onecolumn,twoside]{IEEEtran}
\usepackage{cite}

\ifCLASSINFOpdf
\usepackage[pdftex]{graphicx}
\else
\usepackage[dvips]{graphicx}
\fi
\usepackage{epstopdf}					
\graphicspath{{figures/}}

\usepackage{algorithm}
\usepackage{algorithmic}

\usepackage[cmex10]{amsmath}
\usepackage{amsfonts}

\usepackage[footnotesize,belowskip=-14pt,aboveskip=0pt]{caption}
\usepackage{flushend}
\usepackage{color,soul}
\begin{document}
%
\title{Adaptive Bit Allocation for OFDM Cognitive Radio Systems with Imperfect Channel Estimation}

\author{{Ebrahim Bedeer, 
Mohamed Marey, 
Octavia Dobre, and 
Kareem Baddour \IEEEauthorrefmark{2}}\\
\IEEEauthorblockA{Faculty of Engineering and Applied Science, Memorial University of Newfoundland,
St. John's, NL, Canada\\ 
\IEEEauthorrefmark{2} Communications Research Centre, Ottawa, ON, Canada\\
Email: \{e.bedeer, mmarey, odobre\}@mun.ca, kareem.baddour@crc.ca}
}



\maketitle

\begin{abstract}
Cognitive radios hold tremendous promise for increasing the spectral   
efficiency of wireless communication systems. In this paper, an   
adaptive bit allocation algorithm is presented for orthogonal   
frequency division multiplexing (OFDM) CR systems operating in a   
frequency selective fading environment. The algorithm maximizes the CR  
  system throughput in the presence of narrowband interference, while   
guaranteeing a BER below a predefined threshold. The effect of   
imperfect channel estimation on the algorithm's performance is also   
studied.


\end{abstract}


%
\IEEEpeerreviewmaketitle

\input{intro}

\input{model}
\input{algo_wyg}
\input{sim}
\input{future}
\section*{Acknowledgment}

This work has been supported in part by the Communications Research Centre, Canada.



%



\end{document}

%% file: intro.tex
\section{Introduction}

Cognitive Radio (CR), first introduced by Mitola in \cite{Mitola1999}, is a promising wireless communication paradigm that is aware of its radio surroundings and adapts intelligently to improve spectrum efficiency. Unlicensed or cognitive users (CUs) seek to \textit{underlay}, \textit{overlay}, or \textit{interweave} their signals with licensed or primary users (PUs) \cite{kolodzy2005cognitive, srinivasa2007cognitive, goldsmith2009breaking}. The underlay approach allows concurrent transmission of PUs and CUs as in ultrawide band (UWB) systems. CUs spread their transmission over a wide bandwidth, hence their interference is below an acceptable noise floor to PUs. The overlay approach also allows concurrent transmission of PUs and CUs with a premise that CUs can use part of their power to assist/relay PUs transmission. The interweave approach allows CUs to opportunistically access voids in PUs frequency bands/time slots under the condition that no harmful interference occurs to PUs. In this paper, we focus on the interweave CR systems.

Orthogonal frequency division multiplexing (OFDM) is recognized as an attractive modulation technique for CR due to its flexibility and adaptivity in allocating vacant radio resources among CUs \cite{weiss2004spectrum}. In conventional OFDM-based systems, a fixed bit allocation is used on all subcarriers. Thus, the total BER is dominated by the subcarriers that have the worst performance. To improve the system's BER, adaptive bit allocation can be employed such that the information is redistributed across subcarriers according to the channel state information. Consequently, adaptive bit allocation requires accurate channel estimation at the receiver and a reliable feedback channel between the receiver and the transmitter. 

Different adaptive bit allocation algorithms are presented in the literature \cite{hughes1988ensemble, de1998optimal, levin2001complete, wyglinski2005bit, fox1966discrete,  papandreou2005new, goldsmith1997variable, fischer1996new, kalet1989multitone}. These algorithms can be categorized according to their operation as follows: greedy algorithms \cite{hughes1988ensemble, de1998optimal, levin2001complete, wyglinski2005bit, fox1966discrete,  papandreou2005new}, and water-filling based algorithms \cite{goldsmith1997variable, fischer1996new, kalet1989multitone}.

Greedy algorithms provide optimal performance by incrementally allocating an integer number of bits at the cost of high complexity. This was first suggested by Hughes-Hartog in \cite{hughes1988ensemble}, where one bit is added at a time to the subcarrier requiring the smallest incremental power to maximize the throughput. Unfortunately, the algorithm is complex and it converges very slowly. Campello de Souza \cite{de1998optimal} and Levin \cite{levin2001complete} developed a complete and mathematically verifiable algorithm, known as ``Levin-Campello,'' that offers significant improvement to the work of Hughes-Hartog. 


On the other hand, water-filling based algorithms formulate the adaptive bit allocation problem as a constrained optimization problem that can be solved by classical optimization methods \cite{cover2004elements}. These algorithms allocate non-integer number of bits to subcarriers in a non-iterative manner. Hence, it compromises performance for lower complexity, as it is generally followed by a rounding-off step to allocate an integer number of bits to the transmitted symbols across all subcarriers, thus lowering the overall data rate \cite{papandreou2005new}.

The work in \cite{hughes1988ensemble, de1998optimal, levin2001complete, wyglinski2005bit, fox1966discrete,  papandreou2005new, goldsmith1997variable, fischer1996new, kalet1989multitone} assumes that the OFDM system is interference-free. In this paper, we consider the coexistence between an OFDM CU and a narrowband (NB) PU, and present an adaptive bit allocation algorithm to maximize the average throughput of the CU under an average BER constraint in the presence of the NB interference. The level of NB interference depends on how close the OFDM CU transmits in frequency when compared to the NB PU, for a certain NB PU power. Moreover, the effect of imperfect channel estimation on the algorithm performance is investigated. On the other hand, the OFDM CU spectrum leakage to the NB PU can be reduced by straightforwardly combining the proposed algorithm with spectrum sculpting techniques \cite{budiarjo2006combined,bedeer2011partial}; this is beyond the scope of this paper.

The remainder of the paper is organized as follows. Section \ref{sec:model} presents the system models. Section \ref{sec:algo} delineates the adaptive bit allocation algorithm. Simulation results are presented in Section \ref{sec:sim}. Finally, conclusions are drawn in Section \ref{sec:future}.

%% file: model.tex
\section{System Models} \label{sec:model}

\subsection{OFDM system model}
\begin{figure}[!t]
	\centering
		\includegraphics[width=0.50\textwidth]{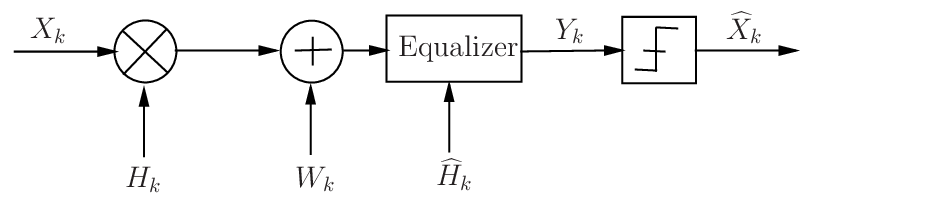}
	\caption{OFDM subcarrier model with imperfect channel estimation.}
	\label{fig:ch}
\end{figure}

We consider an OFDM system with $N$ subcarriers that are orthogonal to each other and intersymbol interference (ISI) free. Hence, each subcarrier can be detected independently using a simple maximum likelihood detector, as shown in Fig.~\ref{fig:ch}. Let $X_k$, $Y_k$, $H_k$, $\widehat{H}_k$, and $W_k$ represent the transmitted symbol, the received symbol, the channel gain, the estimated channel gain, and the additive white Gaussian noise (AWGN), respectively, for subcarrier $k$. The received symbol $Y_k$ can be written as
\begin{eqnarray}
Y_k & =  &\frac{1}{\widehat{H}_k}\Big ( H_k X_k + W_k  \Big ) = X_k \Big ( 1 + \frac{H_k - \widehat{H}_k}{\widehat{H}_k} \Big ) + W'_k,\nonumber \\
 & = & X_k +  \frac{H_k - \widehat{H}_k}{\widehat{H}_k} X_k + W'_k,
\end{eqnarray}
where $\frac{H_k - \widehat{H}_k}{\widehat{H}_k} X_k$ represents the error added to the transmitted symbol on subcarrier $k$ due to imperfect channel estimation. Leke \textit{et al.} in \cite{leke1998impact,wyglinski2005bit} show that this error has a Gaussian distribution with zero mean and variance $\sigma_{h}^2$ given by 
\begin{eqnarray}
\sigma_{h}^2 = \left | \frac{H_k - \widehat{H}_k}{\widehat{H}_k} \right |^2 \sigma_{s,k}^2,
\end{eqnarray}
where $\sigma_{s,k}^2$ is the transmitted symbol power per subcarrier $k$. 

\subsection{NB system model}
We consider a NB PU, coexisting with the OFDM CU, whose signal is given by
\begin{eqnarray}
I(t) & = &\sum_{l =-\infty }^{\infty}b_{l}\:p(t-lT-\xi) \:e^{j2\pi f_{c}t}, 
\end{eqnarray}
where $b_{l}$ is the data symbol transmitted in the $l$th period, $p(t)$ is the impulse response of the transmit filter, $T$ is the symbol period, $\xi$ is the time delay, and $f_{c}$ is the NB PU frequency deviation from the OFDM carrier frequency. 

The instantaneous signal-to-interference-plus-noise ratio (SINR) on subcarrier $k$, $\gamma_k$, in case of imperfect channel estimation and NB interference can be written as
\begin{eqnarray}
\gamma_k = \frac{\sigma_{s,k}^2}{\sigma_{n}^2 + \sigma_{h}^2 + \sigma_{I,k}^2}|H_k|^2,
\label{eq:SINR}
\end{eqnarray}
where $\sigma_n^2$ is the noise variance and $\sigma_{I,k}^2$ is the interference variance per subcarrier $k$. The value of $\sigma_{I,k}^2$ is calculated by evaluating the NB PU signal power per subcarrier at the OFDM receiver FFT output. The resultant NB signal after the OFDM receiver FFT in the $r$th OFDM symbol can be written as 
\begin{eqnarray}
Z^r_k = \frac{1}{\sqrt{N}}\sum_{n=0}^{N-1} \sum_{l =-\infty }^{\infty}b_{l}\: p\Big((n + rN)T_s-lT-\xi\Big) \nonumber \\ \: e^{j2\pi f_{c}(n+rN)T_s} \:e^{-j\frac{2\pi kn}{N}}.
\end{eqnarray}

Hence, the variance of NB signal after the OFDM receiver FFT for the $r$th OFDM block can be written as
\begin{eqnarray}
(\sigma^{r}_{I,k})^2 & = & \mathbb E\left \{ Z^r_k Z^{*r}_k \right \} \nonumber \\
 & = & \frac{\sigma_b^2}{N} \sum_{n=0}^{N-1}\sum_{n'=0}^{N-1} \sum_{l =-\infty }^{\infty} p((n + rN)T_s-lT-\xi) \nonumber \\ & & p((n' + rN)T_s-lT-\xi) \: e^{j2\pi f_{c}(n-n')T_s} \nonumber \\
 & & \: e^{-j\frac{2\pi k}{N}(n-n')},
\end{eqnarray}
where $\mathbb E\{.\}$ denotes the statistical expectation operator, and $\sigma_b^2$ = $\mathbb E\{b_l b^*_l\}$. By averaging over the number of OFDM symbols $R$, the NB PU variance per subcarrier can be finally expressed as
\begin{eqnarray}
 \sigma^2_{I,k} & = & \lim_{R \rightarrow \infty} \frac{1}{R} \frac{\sigma_b^2}{N} \sum_{r = 0}^{R-1} \sum_{n=0}^{N-1}\sum_{n'=0}^{N-1} \sum_{l =-\infty }^{\infty} \nonumber \\
 & & p((n + rN)T_s-lT-\xi) \: p((n' + rN)T_s-lT-\xi) \nonumber \\
 & & e^{j2\pi f_{c}(n-n')T_s} \: e^{-j\frac{2\pi k}{N}(n-n')}.
 \label{eq:sigma}
\end{eqnarray}

%% file: algo_wyg.tex
\section{Adaptive Bit Allocation Algorithm} \label{sec:algo}

This section presents the adaptive bit allocation algorithm to maximize the OFDM CU throughput, i.e. the transmitted number of bits per OFDM symbol, while maintaining the average BER across all OFDM subcarriers below a target BER, in the presence of a NB interference and imperfect channel estimation. This can be formally expressed as
\begin{eqnarray}
\underset{m_k}{max} & & \sum_{k=1}^{N}m_k, \nonumber \\
\textup{subject to} & & \overline{BER} = \frac{\sum_{k=1}^{N}m_k (BER_k)}{\sum_{k=1}^{N}m_k}\leq BER_T,
\end{eqnarray}
where $m_k$ is the number of bits allocated to subcarrier $k$, $\overline{BER}$ is the mean BER, $BER_T$ is the target BER, and $BER_k$ is the BER for subcarrier $k$.

The idea behind the algorithm is to load all subcarriers with the highest possible constellation size and uniform power, and then calculate the BER per subcarrier depending on the channel state condition and the NB interference per subcarrier. The average BER is finally calculated and checked against the target BER. If the average BER meets the target BER, then the final bit allocation is reached; otherwise, the signal constellation on the worst performance subcarrier is decreased and the process repeats. 

The modulation schemes considered in this work are BPSK, QPSK, 16-QAM, and 64-QAM, i.e. each subcarrier can be loaded with a symbol drawn from one of the previously mentioned modulation schemes. The closed-form expression for the $BER_k$ of B/QPSK PSK is given by \cite{proakisdigital}
\begin{eqnarray}
BER_k &{} = &{} Q\left ( \sqrt{2\frac{T_u}{T_o}\gamma_k} \right ), 
\label{eq:PSK}
\end{eqnarray}
where $T_u$ and $T_o$ are the useful OFDM symbol duration and the OFDM symbol duration including the cyclic prefix, respectively, $\frac{T_u}{T_o}$ is the loss due to cyclic prefix, and $Q(.)$ represents the Q-function. The closed-form expression for the $BER_k$ of 16-QAM and 64-QAM is given by \cite{proakisdigital}
\begin{eqnarray}
BER_k & = &\frac{4}{m_k}\left ( 1-\frac{1}{\sqrt{M_k}} \right ) Q\left ( \sqrt{\frac{3}{M_k-1}\frac{T_u}{T_o}\gamma_k} \right )  \nonumber \\
 & &\left ( 1- ( 1-\frac{1}{\sqrt{M_k}} ) Q\left ( \sqrt{\frac{3}{M_k-1}\frac{T_u}{T_o}\gamma_k} \right ) \right ), \nonumber \\
 \label{eq:QAM}
\end{eqnarray}
where $M_k$ is the constellation size per subcarrier $k$. Note that (\ref{eq:PSK}) and (\ref{eq:QAM}) are written under the assumption of Gaussian interference at the OFDM receiver FFT output. As the interference per subcarrier after the OFDM receiver FFT represents the contribution of the interference samples at the FFT input, this tends to have a normal distribution according to the central limit theorem \cite{papoulis1965probability}.

The algorithm can be formally described as follows
\floatname{algorithm}{}
\begin {algorithm}
\renewcommand{\thealgorithm}{} 
\caption{\textbf{Adaptive Bit Allocation Algorithm}}
\begin{algorithmic}[1]
\STATE Initialization: set the modulation scheme of all the subcarriers to 64-QAM.
\STATE Determine $BER_k$, $k$=1,...,$N$, given the SINR $\gamma_k$ values, using (\ref{eq:PSK}) or (\ref{eq:QAM}).
\STATE Compare $\overline{BER}$ with $BER_T$. If $\overline{BER}$ is less than $BER_T$, the current configuration is kept and the algorithm ends.
\STATE Search for the subcarrier with the worst $BER_k$ and reduce the constellation size. If $m_k$ = 1, null the subcarrier (i.e., set $m_k$ = 0). 
\STATE Recompute $BER_k$ of all subcarriers with changed allocations and return to step 3.
\STATE If $BER_T$ cannot be met, the transmission stops.
\end{algorithmic}
\end{algorithm}

We should mention that \cite{wyglinski2005bit} proposes a similar bit allocation algorithm without considering the interference effect, which is crucial in the CR environment.

%% file: sim.tex
\section{Simulation Setup and Results} \label{sec:sim}

\subsection{Simulation setup}
The parameters of the systems considered in this study are provided in Table \ref{tab:SimPar}. A frequency selective fading channel is used for the OFDM CU. The channel impulse response $h(n)$ has a length of $N_{ch}$ = 5 taps, where their components vary independently and are modeled as complex valued Gaussian random variables with zero mean and an exponential power delay profile \cite{morelli2004timing}   
\setlength{\arraycolsep}{0.0em}
\begin{eqnarray}
\mathbb{E}\{\left | h(n) \right |^2\} = \sigma_c^2 \: e^{-n\Xi}, \qquad n = 0, 1, ..., N_{ch}-1,
\end{eqnarray}
where $\sigma_c^2$ is a constant chosen such that the average energy per subcarrier is normalized to unity, i.e. $\mathbb{E}\{\left | H_k \right |^2\}$ = 1, and $\Xi$ represents the decay factor, $\Xi = \frac{1}{5}$. Also, we assume that the frequency selective channel is fixed over a number of OFDM symbols, and a total of $10^4$ channel realization are considered. A root-raised cosine transmit filter is considered for the adjacent NB system, and the time delay $\xi$ is taken as a random variable uniformly distributed between 0 and the NB symbol duration, $T$. The target BER, $BER_T$, is chosen to be $10^{-4}$. The normalized frequency $F_n$, which represents the spectral distance between the OFDM CU and NB PU, is defined as $F_n = \frac{f_c}{BW}$, where $BW$ is the OFDM bandwidth. 
\begin{table}[!t]
  \centering
  \caption{Simulation parameters.}
    \begin{tabular}{rl}
    \hline
    {\bf } & {\bf OFDM-based system} \\ \hline
    { Bandwidth, $BW$} & 1.25 MHz \\
    { Window roll-off factor, $\beta$} & 0 \\
    { Number of subcarriers, $N$} & 128 \\
    { Subcarrier spacing, $\Delta F$} & 9.7656 kHz \\
    { Useful symbol duration, $T_u$} & 102.4 $\mu$sec \\
    { CP duration, $T_{cp}$} & 0.25$T_u$ = 25.6 $\mu$sec \\
    { Postfix duration, $T_{p}$} & 0 \\ 
    { Modulation type} & BPSK, QPSK, 16-QAM, 64-QAM \\ \hline
    {\bf } & {\bf NB system} \\ \hline
    { Bandwidth, $BW_N$} & 15 kHz \\
    { Roll-off factor, $\alpha_N$} & 0.35 \\
    { Modulation type} & QPSK \\ \hline
    \end{tabular}
  \label{tab:SimPar}
\end{table}

\subsection{Simulation results}
\begin{figure}[!t]
	\centering
		\includegraphics[width=0.50\textwidth]{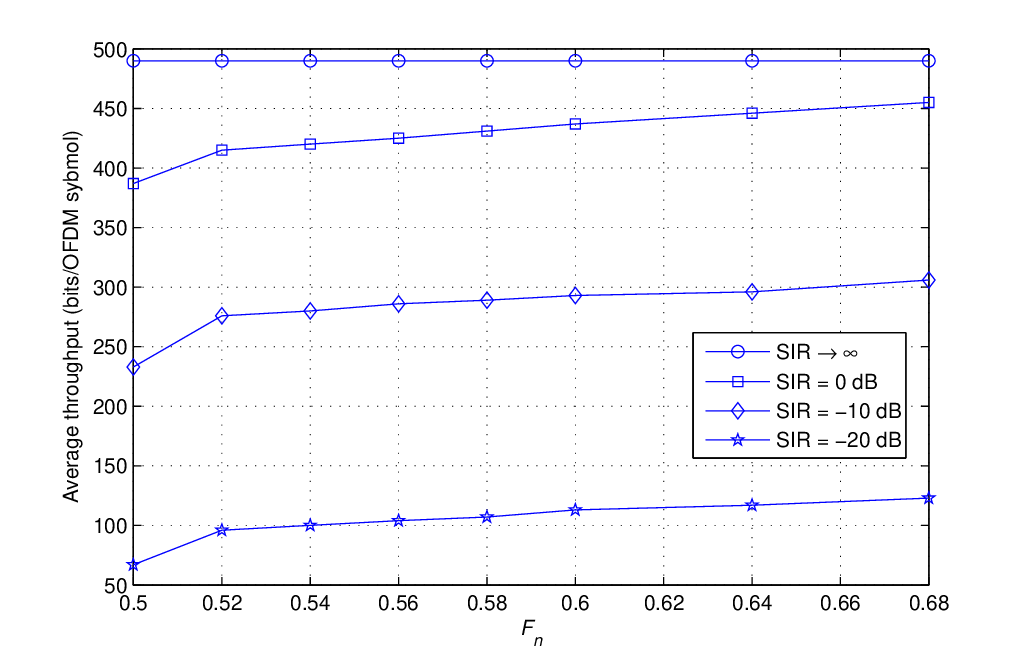}
	\caption{Average throughput as a function of $F_n$ for different SIRs at average SNR = 20 dB in the case of perfect channel estimation.}
	\label{fig:Th_Fn}
\end{figure}

Fig. \ref{fig:Th_Fn} shows the average throughput as a function of $F_n$ for different values of SIRs\footnote[1]{The SIR is defined after the OFDM receiver FFT as the ratio between the OFDM and NB interference average powers, respectively, as in \cite{bedeer2011partial}.} at average SNR\footnote[2]{The average SNR is calculated by averaging the instantaneous SNR values per subcarrier over the total number of subcarriers and the total number of channel realizations, respectively.} = 20 dB, assuming perfect channel estimation, i.e. $\sigma_h^2$ = 0. As one can notice, as $F_n$ increases, the NB PU effect on the OFDM CU decreases; hence, more bits can be transmitted on average per each OFDM symbol while achieving a target BER of $10^{-4}$. Furthermore, the average throughput increases as the SIR increases (i.e. the effect of the NB PU interference on the OFDM subcarriers decreases).

\begin{figure}[!t]
	\centering
		\includegraphics[width=0.50\textwidth]{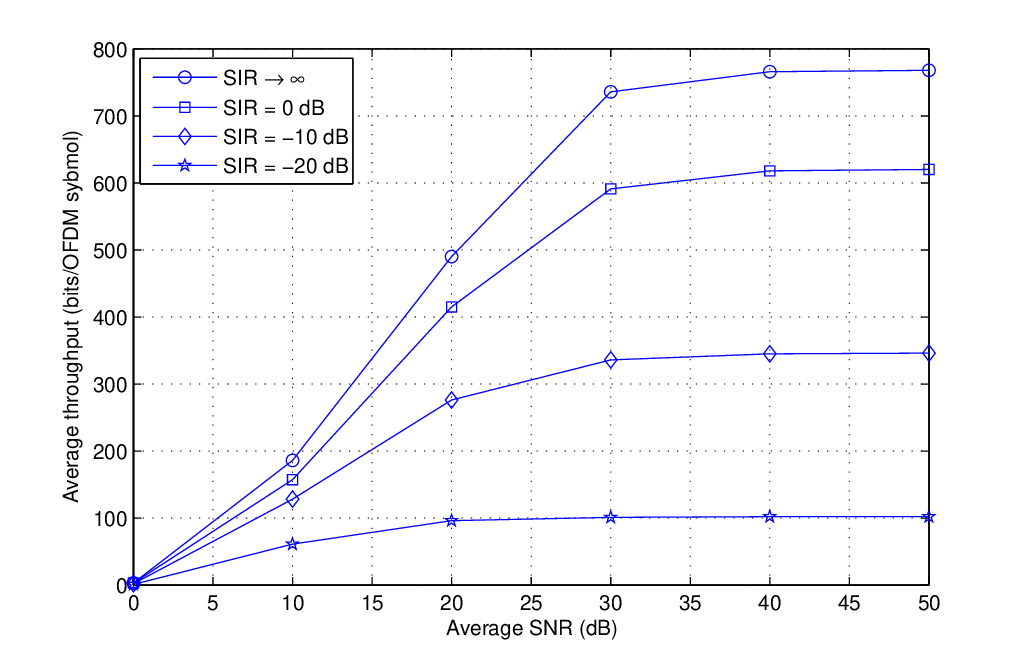}
	\caption{Average throughput as a function of average SNR for different SIRs at $F_n$ = 0.52 in case of perfect channel estimation.}
	\label{fig:Th_SNR}
\end{figure}

Fig. \ref{fig:Th_SNR} depicts the average throughput as a function of average SNR for different values of SIRs at $F_n$ = 0.52, assuming perfect channel estimation. As one can observe, as the average SNR or SIR increases, more bits can be loaded on OFDM subcarriers while achieving the target BER, which translates into an increase in the average throughput. However, the throughput saturates beyond a certain average SNR at a given SIR. This can be explained as, at a certain average SNR value, the OFDM CU subcarriers are loaded with the maximum constellation given a certain SIR value and a further increase in the average SNR will not improve the average throughput. Moreover, as the SIR increases, the effect of the NB PU on the OFDM CU reduces, which translates into an increase in the average throughput. It is worth pointing out that for SIR values of $- 10$ and $- 20$ dB, the OFDM CU subcarriers adjacent to the NB PU are nulled, which in turn reduces the OFDM CU spectrum leakage to the NB PU.

\begin{figure}[!t]
	\centering
		\includegraphics[width=0.50\textwidth]{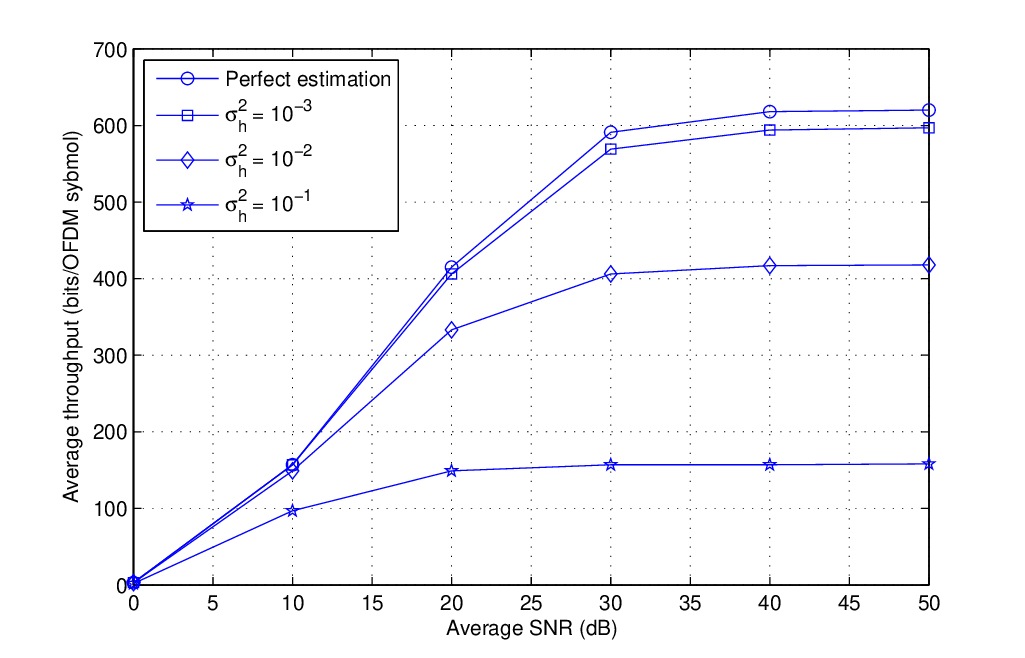}
	\caption{Average throughput as a function of average SNR for different values of $\sigma_h^2$ at $F_n$ = 0.52 and SIR = 0 dB.}
	\label{fig:Th_ch}
\end{figure}

The effect of imperfect channel estimation on the average throughput is shown in Fig. \ref{fig:Th_ch}. The average throughput is plotted as a function of average SNR for different values of $\sigma_h^2$ for $F_n$ = 0.52 and SIR = 0 dB. As expected, increasing the channel estimation error variance, $\sigma_h^2$, reduces the average throughput. According to (\ref{eq:SINR}), the channel estimation error has the same effect as the interference which can be noticed from Figs. \ref{fig:Th_SNR} and \ref{fig:Th_ch}.

%% file: future.tex
\section{Conclusion} \label{sec:future}

In this paper, we presented an adaptive bit allocation algorithm to maximize the OFDM CU throughput in the presence of NB PU for CR systems with imperfect channel estimation. The average throughput of the OFDM CU is maximized under the constraint of a BER below a target value. As expected, the average throughput increases as the CU-PU frequency separation and SIR increase, respectively. Moreover, increasing the channel estimation error variance reduces the average throughput. In future work, we will extend the proposed algorithm to include power loading, and impose a constraint on the maximum transmit power to reduce the OFDM CU spectrum leakage to PUs, in addition to considering spectrum sculpting techniques. 